\documentclass[aps,twocolumn,a4paper,pra,superscriptaddress,floatfix]{revtex4}%

\usepackage{amssymb}
\usepackage{amsmath}
\usepackage{amsfonts}
\usepackage{graphicx}
\usepackage{graphics}

\def\G{\mathrm{G}}

\def\Cas{\mathrm{Cas}}
\def\perf{\mathrm{perf}}
\def\plas{\mathrm{plas}}
\def\Drud{\mathrm{Drud}}
\def\T{\mathrm{T}}
\def\B{\mathrm{B}}

\def\bk{\mathbf{k}}

\def\dd{\mathrm{d}}

\def\TE{\mathrm{TE}}
\def\TM{\mathrm{TM}}

\def\P{\mathrm{P}}    
\def\C{\mathrm{C}}    
\def\GC{G_\C} 

\def\kP{k_\mathrm{P}}    
\def\kC{k_\C}    

\def\perf{\mathrm{perf}}
\def\PFA{\mathrm{PFA}}
\def\Tr{\mathrm{Tr}}
\def\calF{\mathcal{F}}
\def\calE{{E}}

\def\calR{{R}}
\def\calK{{K}}
\def\calD{{D}}
\def\P{\mathrm{P}}    
\def\P{\mathrm{P}}    
\def\S{\mathrm{S}}    
\def\PFA{\mathrm{PFA}}
\def\TE{\mathrm{TE}}
\def\TM{\mathrm{TM}}
\def\bk{\mathbf{k}}
\def\dd{\mathrm{d}}

\def\G{\mathrm{G}}

\def\cF{\mathcal{F}}

\def\cL{\mathcal{L}}
\def\cK{{K}}
\def\cR{{R}}

\def\max{\mathrm{max}}

\def\lone{{\ell=1}}

\begin{document}

\title{Casimir effect in the scattering approach: correlations between material properties, temperature and geometry}

\author{Astrid Lambrecht}
\author{Antoine Canaguier-Durand}
\author{Romain  Gu\'erout} 
\author{Serge Reynaud}
\affiliation{Laboratoire Kastler Brossel,                  
CNRS, ENS, Universit\'e Pierre et Marie Curie case 74,
Campus Jussieu, F-75252 Paris Cedex 05, France}

\date{\today}

\begin{abstract}                                                                     
We present calculations of the quantum and thermal Casimir interaction
between real mirrors in electromagnetic fields using the scattering approach.
We begin with a pedagogical introduction of this approach in simple cases
where the scattering is specular. We then discuss the more general case of
stationary arbitrarily shaped mirrors and present in particular applications
to two geometries of interest for experiments, that is corrugated plates
and the plane-sphere geometry. The results nicely illustrate the rich
correlations existing between material properties, temperature and
geometry in the Casimir effect.
\end{abstract}

\maketitle

\section{Introduction}

The Casimir effect \cite{Casimir}
is an observable effect of vacuum fluctuations in the mesoscopic world,
to be tested with the greatest care as a crucial prediction of quantum field theory
\cite{Milonni94,LamoreauxResource99,Reynaud01,Bordag01,DeccaAP05,Milton05,LambrechtNJP06}.
It also constitutes a fascinating interface between quantum field theory and
other important aspects of fundamental physics, for example through
its connection with the problem of vacuum energy \cite{Jaekel97,Elizalde07,Jaekel08}.

Casimir physics plays an important role in the tests of gravity
at sub-millimeter ranges \cite{Fischbach98,Adelberger03}.
Strong constraints have been obtained in short range
Cavendish-like experiments \cite{Kapner07}.
A hypothetical new force of Yukawa-like form could not exceed
the gravitational force in the range above 56$\mu$m.
For ranges of the order of the micrometer, similar tests are performed
by comparing the results of Casimir force measurements with theoretical predictions
\cite{LambrechtPoincare,OnofrioNJP06,DeccaEPJ07}. At even shorter scales,
those tests can be performed using atomic \cite{Lepoutre09} or nuclear
\cite{Nesvizhevsky08} force measurements.
In any of these short-range gravity tests, a new hypothetical
force would appear as a difference between the experimental result
$F_\mathrm{exp}$ and the theoretical prediction $F_\mathrm{th}$.
This implies that $F_\mathrm{th}$ and $F_\mathrm{exp}$ have to be assessed independently from each other
and necessarily forbids to use the theory-experiment comparison
for proving (or disproving) some specific experimental result or theoretical model.

Finally, the Casimir force and the closely related Van der Waals force
are dominant at micron or sub-micron distances, entailing their
strong connections with various important domains,
such as atomic and molecular physics, condensed matter and surface physics,
chemical and biological physics, micro- and nano-technology
\cite{Parsegian06}.

\section{Comparison of Casimir force measurements with theory}
\label{Comp}

Casimir calculated the force between a pair of perfectly smooth, flat and
parallel plates in the limit of zero temperature and perfect reflection which led him
to the universal expressions for the force $F_\Cas$ and energy $E_\Cas$
\begin{equation}
F_\Cas=-\frac{\hbar c \pi ^2 A}{240L^4} \quad,\quad
E_\Cas= - \frac{\hbar c \pi^2 A}{720 L^3}.
\label{FECasimir}
\end{equation}
with $L$ the mirrors' separation, $A$ their surface,
$c$ the speed of light and $\hbar$ the Planck constant.
The universality of these ideal Casimir formulas is explained by the saturation
of the optical response of perfect mirrors which exactly reflect 100\% of the incoming fields.
This idealization does not correspond to any real mirror.
In fact, the effect of imperfect reflection is large in most experiments, and
a precise knowledge of its frequency dependence is essential for obtaining
reliable theoretical predictions to be compared with Casimir force measurements
\cite{Sparnaay89,Lamoreaux97,Mohideen98,Harris00,Ederth00,Bressi02,Decca03prl,ChenPRA04,DeccaPRD07,Munday07,vanZwol08,Munday09,Jourdan09,Masuda09,deMan09}.

\subsection{The description of metallic mirrors}

The most precise experiments are performed with metallic mirrors which are
good reflectors at frequencies smaller than their plasma frequency $\omega_\P$.
Their optical response at a frequency $\omega$ is described by a reduced dielectric function
written as
\begin{equation}
\varepsilon \left[\omega\right] = \bar{\varepsilon}\left[\omega\right] +
\frac{\sigma \left[\omega\right] }{-i\omega}
\quad,\quad \sigma \left[\omega\right] = \frac{\omega_\P^2}{\gamma-i\omega} .
\label{Drude}
\end{equation}
The function $\bar{\varepsilon} \left[\omega\right] $ represents the contribution
of interband transitions and it is regular at the limit $\omega\to0$.
Meanwhile $\sigma \left[\omega\right]$ is the reduced conductivity,
measured as a frequency (the SI conductivity is $\epsilon_0\sigma$),
which describes the contribution of the conduction electrons.

A simplified description corresponds to the lossless limit $\gamma \to 0$
often called the plasma model. As $\gamma$ is much smaller than
$\omega_\P$ for good conductors, this simple model captures
the main effect of imperfect reflection. However it cannot be considered
as an accurate description since a much better fit of tabulated optical data
is obtained with a non null value of $\gamma$ \cite{LambrechtEPJ00,SvetovoyPRB08}.
Furthermore, the Drude model, with $\gamma\neq0$, meets the important property
of ordinary metals which have a finite static conductivity
$\sigma_0 = \frac{\omega_\P^2}{\gamma}$,
in contrast to the lossless limit which corresponds to an infinite value
for $\sigma_0$.

Another correction to the Casimir expressions is associated with the effect
of thermal fluctuations \cite{Mehra67,Brown69,Schwinger78,GenetPRA00}.
Bostr\"om and Sernelius have remarked that the small non zero value of $\gamma$
had a significant effect on the force evaluation at ambient
temperature \cite{Bostrom00}. This significant difference is attributed to the
vanishing contribution of TE modes at zero frequency for dissipative
mirrors entailing that for the Casimir force, contrary to the dielectric function,
there is no continuity from the Drude to the plasma model at the limit of a
vanishing relaxation.
The ratio between the predictions evaluated at $\gamma=0$ and $\gamma\neq0$
even reaches a factor 2 at the limit of large temperatures or large distances.
Unfortunately it has not yet been possible to test this striking prediction
since the current experiments do not explore this domain.

The current status of Casimir experiments appears to favor theoretical predictions
obtained with the lossless plasma model $\gamma=0$ rather than those corresponding 
to the Drude model with $\gamma\neq0$ as one might have expected (see Fig.1 in \cite{DeccaPRD07}).
We thus have to face a discrepancy between theory and experiment.
This discrepancy may have various origins, in particular
artefacts in the experiments or inaccuracies in the calculations.
They may also come from yet unmastered differences between the situations
studied in theory and the experimental realizations.

These remarks have led to a blossoming of papers devoted to the thermal effect
on the Casimir force, for reviews see \textit{e.g.}
\cite{Reynaud03,Klim06,BrevikNJP06,BrevikJPA08,Milton08R}.
It is worth emphasizing that microscopic descriptions
of the Casimir interaction between two metallic bulks lead to predictions
agreeing with the lossy Drude model rather than the lossless plasma model
at the limit of large temperatures or large distances
\cite{Jancovici05,Buenzli05,Bimonte09}.

It is also important to note that the Drude model leads to a negative contribution
of the Casimir interaction to entropy, in contrast to the plasma model \cite{Bezerra02}.
There is no principle inconsistency with the laws of thermodynamics at this point since the negative contribution
is nothing but a difference of entropies (see for example \cite{IngoldPRE09}).

\subsection{The role of geometry}

The geometry plays an important role in the context
of theory/experiment comparison for Casimir forces.
Precise experiments are indeed performed between a plane and a sphere
whereas most exact calculations are devoted to the geometry of two parallel plates.
The estimation of the force in the plane-sphere geometry thus involves the so-called
\textit{Proximity Force Approximation} (PFA) \cite{Derjaguin68} which amounts to
averaging the force calculated in the parallel-plates geometry over the distribution
of local inter-plate distances, the force being deduced from the Lifshitz formula
\cite{Lifshitz56,DLP61}, the meaning of which will be discussed below.

This trivial treatment of geometry cannot reproduce the rich interconnection
expected to take place between the Casimir effect and geometry
\cite{Balian,Balian03,Balian04}.
In the plane-sphere geometry in particular, the PFA can only be valid when the
radius $R$ is much larger than the separation $L$ \cite{Schaden00,Jaffe04,JaffePRA05}.
But even if this limit is met in experiments, the PFA gives no information about
its accuracy for a given ratio of $L/R$ and how this accuracy depends on the
properties of the mirror, on the distance or temperature.

Answers to these questions can only be obtained by pushing the theory beyond
the PFA, which has been done in the past few years
\cite{ReynaudJPA08,EmigJPA08,BordagJPA08,WirzbaJPA08,KlingmullerJPA08}.
A multipolar expansion of the Casimir effect between perfect mirrors
in electromagnetic vacuum was proposed in \cite{Emig08,Maia08}.
These calculations have now been performed for plane and spherical metallic surfaces
coupled to electromagnetic vacuum, at zero \cite{CanaguierPRL09}
or non zero temperature \cite{CanaguierPRL10,CanaguierPRAsubmitted}, which
has opened the way to a comparison with theory of the only experimental
study devoted to a test of PFA in the plane-sphere geometry \cite{Krause07}.
As we will see at the end of this article, the features of the thermal Casimir force mentioned in section \ref{Comp}
are considerably altered when the geometry is properly taken into account.
The factor of 2 between the force values within Drude and plasma model is reduced to a factor
of 3/2 decreasing even more below this value when small spheres are considered.
Negative entropies are not only found for the Drude model but also for perfect reflector
and plasma models, which means that negative contributions of the Casimir interaction to
entropy can be found even in the absence of dissipation.

Another specific geometry of great interest, that we will present in the following,
is that of surfaces with periodic corrugations. As lateral translation symmetry is broken,
the Casimir force contains a lateral component which is smaller than the normal one,
but has nevertheless been measured in dedicated experiments \cite{Chen02}.
Calculations beyond the PFA have first been performed with the simplifying
assumptions of perfect reflection \cite{Emig05} or shallow corrugations
\cite{Rodrigues06,Rodrigues07}.
As expected, the PFA was found to be accurate only in the limit
of large corrugation wavelengths.
Very recently, experiments have been able to probe the beyond-PFA regime
\cite{Chan08,Chiu09} and exact calculations of the forces between real mirrors
with deep corrugations \cite{Lambrecht08,Lambrecht08b} have been performed.
More discussions on these topics will be presented below.

\section{The scattering approach}

In the following, we will focus our attention onto the scattering approach,
which is an efficient and elegant method for addressing the aforementioned questions.

This method has been used for years for describing the optical properties of
non perfectly reflecting mirrors in terms of scattering amplitudes \cite{Jaekel91,GenetPRA03}.
These scattering amplitudes are often deduced from Fresnel reflection amplitudes calculated
for mirrors described by local dielectric response functions, in which case the expression
of the Casimir force is reduced to the Lifshitz expression \cite{Lifshitz56,DLP61}.
However the scattering approach is much more
general than the Lifshitz one since real mirrors are always described by
some scattering amplitudes but not necessarily by local
dielectric response functions. This point will be discussed in more detail below.

The interest in the scattering approach has considerably increased since it has become clear that it is also an extremely efficient method for
calculating the Casimir effect in non trivial geometries.
This was realized by several groups employing different theoretical techniques
and using different notations (see \cite{Milton08} for an historial overview).
Besides the already quoted papers, one may cite the following references
which used different versions of the scattering approach
\cite{Bulgac06,Bordag06,Kenneth06,Emig07,Rahi09}
or alternative methods \cite{Gies03,Gies06,Emig06,Dalvit06a,Dalvit06b,Rodriguez07}.
This topic has seen recently an impressive number of new applications
proposed, among ones one may cite
\cite{Dobrich2009,Bordag2009,Gies2009,Bordag2010,Emig2010,Zandi2010,Weber2010a,Weber2010b}.

The first explicit application of the scattering approach to non-trivial geometries
and non perfect reflectors was developed in \cite{Maia05,MaiaNeto05} to calculate the
roughness correction to the Casimir force between two planes, in a perturbative
expansion with respect to the roughness amplitude.
The same perturbative formalism was also applied to compute the lateral Casimir force
\cite{Rodrigues06,Rodrigues07} and the Casimir torque \cite{RodriguesEPL06}
between two corrugated surfaces made of real material, and then
to derive the Casimir-Polder potential for an atom near a corrugated surface \cite{Dalvit08,Messina09}.

Let us recall that results applicable to the non retarded case have been available
\cite{Johannson97,Noguez04} before those corresponding to the full retarded theory,
and also that the scattering theory has been used for a long time for studying
the Casimir-Polder force between atoms or molecules \cite{Feinberg70,Power94}.

We begin the review of the scattering approach by an introduction
considering the two simple cases of the Casimir force between two scatterers
on a 1-dimensional line and between two parallel plates coupled
through specular scattering to 3-dimensional electromagnetic fields
\cite{Jaekel91}. We then address the general case of non specular scatterers
in 3-dimensional electromagnetic fields \cite{LambrechtNJP06}.

\subsection{Mirrors on a 1-dimensional line}

The first case corresponds to the quantum field theory in 2-dimensional spacetime
(1-d space plus time). In this simple case, we have to consider only two scalar fields
counter-propagating along opposite directions.
The results summarized below are drawn from a series of papers devoted to the study of
static or dynamic Casimir force between mirrors coupled to these scalar fields
\cite{Jaekel97,Jaekel91,QO92,JdP92,PLA92,PLA93a,JdP93a,JdP93c,PLA93b,PRL96b,PLA97}.
For example, it was established in \cite{JdP93c} that the Casimir energy
does contribute to the inertia of the cavity as it should according to the principles
of relativity.

In this simple model, a mirror M$_1$ at rest at position $q_1$ is described by a 2x2 scattering
matrix $S_1$ containing reflection and transmission amplitudes $r_1$ and $t_1$
\begin{eqnarray}
S_1 =\left[
\begin{array}{cc}
t_1 & r_1 e^{-2i\omega q_1/c } \\
r_1 e^{2i\omega q_1/c } & t_1
\end{array}
\right] .
\end{eqnarray}
Two mirrors M$_1$ and M$_2$ at rest at positions $q_1$ and $q_2$
form a Fabry-Perot cavity described by a global scattering matrix $S_{12}$ which can be deduced
from the elementary matrices $S_1$ and $S_2$ associated with the two mirrors.
\begin{eqnarray}
S_{12} = \frac 1d
\left[
\begin{array}{cc}
t_1 t_2 &
d r_2 e^{-i \frac{\omega L}{c}}+
t_2^2 r_1 e^{i \frac{\omega L}{c}} \\
d r_1 e^{-i\ \frac{\omega L}{c}}+
t_1^2 r_2 e^{i \frac{\omega L}{c}} &
t_1 t_2
\end{array}
\right] .
\end{eqnarray}
The denominator $d$ is given by
\begin{eqnarray}
d = 1-r_1 r_2 e^{2i\omega L/c}
\quad,\quad L\equiv q_2-q_1 ,
\end{eqnarray}
and its zeros (the poles of $S_{12}$) represent the resonances of the cavity.
It turns out that the forthcoming discussions of the Casimir effect depend only
on the expression of $d$ and not on all the other details in the form of $S_{12}$.
The reason explaining this property is the following relation between the
determinants of the $S-$matrices (all supposed to be unitary in the simple model)
\begin{eqnarray}
\det S_{12} =  \left(\det S_1\right) \left(\det S_2\right) \left(\frac{d^*}d \right) .
\end{eqnarray}

From this relation, it is easy to derive the Casimir free energy as a variation
of field energy (vacuum energy at $T=0$, vacuum plus thermal energy otherwise).
The presence of a scatterer indeed shifts the field modes and thus induces
a variation of the global field energy.
The Casimir free energy is then obtained as the variation of field energy in presence
of the cavity corrected by the effects of each mirror taken separately \cite{Jaekel91}
\begin{eqnarray}
\label{Fphaseshift}
\mathcal{F} &\equiv& \delta \mathcal{F} _\mathrm{field,12}
-\delta \mathcal{F} _\mathrm{field,1}-\delta \mathcal{F} _\mathrm{field,2} \nonumber \\
&=& - \int_0^\infty \frac{{\rm d}\omega}{2\pi} N \hbar\Delta .
\end{eqnarray}
$\Delta$ is a function of the frequency $\omega$ representing the phaseshift produced by
the Fabry-Perot cavity, again corrected by the effects of each mirror taken separately
\begin{eqnarray}
\Delta(\omega) &  = & \frac{\ln\mathrm{det} S_{12} - \ln\mathrm{det} S_1
- \ln\mathrm{det} S_2}i \nonumber \\
&=& \frac1i \ln\left(\frac{d^*}{d}\right) .
\end{eqnarray}
$N$ is the mean number of thermal photons per mode,
given by the Planck law, augmented by the term $\frac12 $
which represents the contribution of vacuum
\begin{eqnarray}
N(\omega) = \frac12
+ \frac1{\exp\frac{\hbar\omega}{k_\mathrm{B}T}-1}
= \frac1{2 \tanh\frac{\hbar\omega}{2k_\mathrm{B}T}} .
\end{eqnarray}

This phaseshift formula can be given alternative interpretations \cite{Jaekel91}.
In particular, when the Casimir force $F$ is derived from the free energy
\begin{eqnarray}
F&=&-\frac{\partial\mathcal{F}(L,T)}{\partial L}
= \int_0^\infty
\frac{\mathrm{d}\omega}{\pi}
\frac{N \hbar \omega}c \left(f + f^*\right) \nonumber \\
&=& \int_0^\infty
\frac{\mathrm{d}\omega}{\pi}
\frac{N \hbar \omega}c \left(g-1\right) , \\
f &\equiv& \frac{r e^{2i\omega L/c}}
{1-r e^{2i\omega L/c}}  \quad,\quad
g \equiv \frac{1- \left| r e^{2i\omega L/c}\right| ^2}
{\left|1-r e^{2i\omega L/c}\right| ^2} ,
\nonumber
\end{eqnarray}
it is seen as resulting from the difference of radiation pressures exerted
onto the inner and outer sides of the mirrors by the field fluctuations.
For each field mode at frequency $\omega$, $\frac{N \hbar \omega}c$
represents the field momentum while $g$ is the ratio of fluctuation energies
inside and outside the Fabry-Perot cavity.

Using the analytic properties of the causal function $\ln d$,
the Casimir free energy can also be written as an
integral over imaginary frequencies $\omega=i\xi$ (Wick rotation)
\begin{eqnarray}
\mathcal{F} =
\hbar\int\frac{\dd\xi}{2\pi} \cot\left(\frac{\hbar\xi}{2k_\B T}\right) \ln d(i\xi).
\end{eqnarray}
Using the pole decomposition of the cotangent function and the analytic
properties of $\ln d$, this expression can finally be written as a sum
over Matsubara frequencies
\begin{eqnarray}
\mathcal{F} = k_\B T \sum_m{}^\prime \ln d(i\xi_m)
\quad,\quad \xi_m \equiv \frac{2\pi m k_\B T}\hbar.
\end{eqnarray}
The Matsubara sum $\sum_m^\prime$ is the sum over positive integers $m$
with $m=0$ counted with a weight $\frac12$.

For completeness, let us recall also that the contribution to entropy
of the Casimir interaction is simply written as
\begin{eqnarray}
\mathcal{S} \equiv -\frac{\partial\mathcal{F} (L,T)}{\partial T} .
\end{eqnarray}
Hence, it is defined as a difference of entropies just as the
free energy $\mathcal{F}$ has been defined in (\ref{Fphaseshift}) above as a difference of free energies.

\subsection{Specular reflection in 3-d space}

\begin{figure}[h]
\centering
\includegraphics[width=0.25\textwidth]{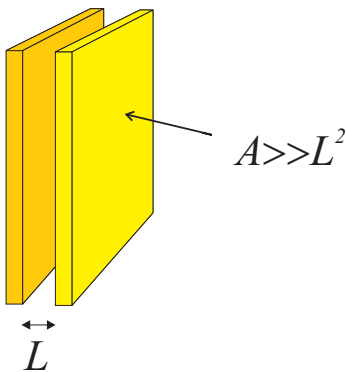}
\caption{Two plane parallel plates at distance $L$ facing each other constitute the Casimir cavity.}
\label{cavity}
\end{figure}

The same lines of reasoning can be followed when studying the case of
two specularly reflecting mirrors coupled to electromagnetic fields
in 3-dimensional space. The geometry is sketched in Fig. \ref{cavity}
with two plane parallel mirrors aligned along the transverse directions
$x$ and $y$ (longitudinal direction denoted by $z$)

Due to the symmetry of this configuration,
the frequency $\omega$, the transverse vector $\bk \equiv \left( k_x,k_y\right)$ and
the polarization $p=\TE,\TM$ are preserved by all scattering processes.
The mirrors are described by reflection and transmission amplitudes
which depend on these scattering parameters.
We assume thermal equilibrium for the whole ``cavity + fields'' system,
and proceed with the derivation as in the simpler case of a 1-dimensional space.
Some elements have to be treated with greater care now \cite{GenetPRA03,LambrechtNJP06}.
First there is a contribution of evanescent waves besides that of ordinary modes
freely propagating outside and inside the cavity and it has to be taken
carefully into account.
The properties of the evanescent waves are described
through an analytical continuation of those of ordinary ones, using the well
defined analytic behavior of the scattering amplitudes.
Then dissipation inside the mirrors may also play a role which implies considering
the additional fluctuation lines coming along with dissipation \cite{LambrechtNJP06,GenetPRA03}.

At the end of this derivation the free energy
may still be written as a Matsubara sum
\begin{eqnarray}
\label{CasimirFreeEnergy}
&&\calF =
k_\B T \sum_\bk \sum_p \sum_m{}^\prime \ln d(i\xi_m,\bk,p)
\quad,\\
&& \xi_m \equiv \frac{2\pi m k_\B T}\hbar \quad, \quad \sum_\bk \equiv A \int\frac{\dd^2\bk}{4\pi^2} \equiv A \int\frac{\dd k_x\dd k_y}{4\pi^2} . \nonumber
\end{eqnarray}
$\sum_\bk$ is the sum over transverse wavevectors with $A$ the area of the plates,
$\sum_p$ the sum over polarizations and $\sum_m{}^\prime$ the same Matsubara sum as in the 1-d case.
The denominator is now written in terms of the result $\kappa$ of Wick rotation
on the longitudinal wavevector $k_z$
\begin{eqnarray}
d(i\xi,\bk,p) &=& 1 - r_1(i\xi,\bk,p) r_2(i\xi,\bk,p)
\exp^{ -2\kappa L } , \\
\kappa &\equiv& \sqrt{\bk^2+\frac{\xi^2}{c^2}} . \nonumber
\end{eqnarray}

This expression reproduces the ideal Casimir formula (\ref{FECasimir})
in the limits of perfect reflection $r_1 r_2 \rightarrow 1$
and zero temperature $T \rightarrow 0$.
It is valid and regular at thermal equilibrium at any temperature
and for any optical model of mirrors obeying causality and high frequency
transparency properties \cite{LambrechtNJP06,Jaekel91,GenetPRA03}.
It can thus be used for calculating the Casimir force between arbitrary
mirrors, as soon as the reflection amplitudes are specified.
These amplitudes are commonly deduced from models of mirrors,
the simplest of which is the well known Lifshitz model
\cite{Lifshitz56,DLP61} which
corresponds to semi-infinite bulk mirrors characterized by a
local dielectric response function $\varepsilon (\omega)$
and reflection amplitudes deduced from the Fresnel law
\begin{eqnarray}
\label{Fresnel}
&&r_\TE(\bk,\xi) = \frac{\kappa-\kappa_t}{\kappa+\kappa_t}
\quad,\quad
r_\TM(\bk,\xi)=\frac{\epsilon \kappa-\kappa _{t}}{%
\varepsilon \kappa +\kappa _{t}} ,
\\
&&\kappa _t\equiv\sqrt{\mathbf{k}^{2}+\varepsilon \,\frac{{\xi }^{2}%
}{c^{2}}} . 
\end{eqnarray}
$\varepsilon$ is the dielectric function (\ref{Drude}) and $\kappa_t$
denotes the result of Wick rotation of the longitudinal wavevector
inside the medium.

In the most general case, the optical response of the mirrors
cannot be described by a local dielectric response function.
The expression (\ref{CasimirFreeEnergy}) of the free energy is still valid
in this case with the reflection amplitudes to be determined from
microscopic models of mirrors.
Recent attempts in this direction can be found for example in
\cite{Pitaevskii08,Geyer09,Pitaevskii09r,Dalvit08b,Decca09,Dalvit09r,Svetovoy08}.

At this stage, several remarks can be addressed to the readers interested in historical details:
\begin{itemize}
\item The Lifshitz expression was not written in terms of reflection amplitudes
until Kats noticed that this formulation was natural \cite{Kats77}.
To our best knowledge, the first appearance of an expression of the Casimir effect
in terms of reflection amplitudes corresponding to an arbitrary microscopic model
(not necessarily a dielectric response function) is in \cite{Jaekel91}.

\item The fact that the expression (\ref{CasimirFreeEnergy}) of the free energy
is valid for lossy as well as lossless mirrors is far from obvious.
In the lossy case, one has indeed to take into account the contributions of
fluctuations coming from the additional modes associated with dissipation.
This property has been demonstrated with an increasing range of validity
in \cite{Jaekel91}, \cite{GenetPRA03} and \cite{LambrechtNJP06}
(see also \cite{Barnett98} for a theorem playing a crucial role
in this demonstration).

\item
The question had been asked in \cite{Reynaud03} whether the regularity conditions
needed to write the Matsubara sum were met for the Drude model which
shows discontinuities at $\xi\to0$. This question
has been answered positively in \cite{IngoldPRE09}.

\end{itemize}

\subsection{The non-specular scattering formula}

We now present a more general scattering formula allowing one to calculate
the Casimir force between stationary objects with arbitrary shapes. We
restrict our attention to the case of disjoint objects, exterior to each
other, which corresponds to the configuration initially considered
by Casimir (for interior configurations, which may be treated with
similar techniques, see for example
\cite{Dalvit04,Mazzitelli06,BordagNikolaev09,Zaheer09}).

The main generalization with respect to the already discussed specular cases is that
the scattering matrix $\mathbb{S}$ has now to account for non-specular reflection.
It is therefore a much larger matrix which mixes different wavevectors and polarizations
while preserving frequency as long as the scatterers are stationary \cite{LambrechtNJP06}.
Of course, the non-specular scattering formula is the generic one
while specular reflection can only be an idealization.

As previously, the Casimir free energy can be written as the sum
of all the phaseshifts contained in the scattering matrix
\begin{eqnarray}
\calF& =& i\hbar \int_0^\infty \frac{\dd \omega}{2\pi}
N(\omega) \ln \det \mathbb{S} \nonumber \\
&=& i\hbar \int_0^\infty \frac{\dd \omega}{2\pi}
N(\omega) \Tr \ln \mathbb{S} . 
\end{eqnarray}
The symbols $\det$ and $\Tr$ refer to determinant and trace over the modes of
the scattering matrix at a given frequency $\omega$.
After a Wick rotation the formula can still be written as a Matsubara sum
\begin{eqnarray}
\label{CasimirFreeEnergyNS}
\calF = k_\B T \sum_m{}^\prime \Tr \ln \calD (i\xi_m)
\quad,\quad \xi_m \equiv \frac{2\pi m k_\B T}\hbar .
\end{eqnarray}
The matrix $\calD$ (here written at Matsubara frequencies $\omega_m=i\xi_m$)
is the denominator of the scattering matrix. It describes the resonance
properties of the cavity formed by the two objects 1 and 2 and may
be written as
\begin{eqnarray}
\calD= 1 - \calR_1 \exp^{ -\calK L } \calR_2 \exp^{ -\calK L }  . &&
\end{eqnarray}
The matrices $\calR_1$ and $\calR_2$ represent reflection on the two objects
1 and 2 respectively while $\exp^{-\calK L}$ describes propagation
inbetween reflections on the two objects.
Note that the matrices $\calD$, $\calR_1$ and $\calR_2$, which were diagonal
in the plane wave basis for specular scattering,
are no longer diagonal in the general case of non specular scattering.
The propagation factors remain diagonal in this basis with their eigenvalues $\kappa$
written as in (\ref{CasimirFreeEnergy}). Clearly the expression
(\ref{CasimirFreeEnergyNS}) does not depend on the choice of a specific basis.
We remark also that (\ref{CasimirFreeEnergyNS}) takes a simpler form in the limit
of zero temperature
\begin{eqnarray}
&&F=-\frac{\dd \calE}{\dd L} \quad,\quad
\calE = \hbar \int_0^\infty \frac{\dd \xi}{2\pi} \ln \det \calD (i\xi) .
\label{CasimirEnergyTnull}
\end{eqnarray}
Applications to be presented in the next sections will also involve the Casimir force
gradient $G$ which is often measured in experiments and defined as
\begin{equation}
G = - \frac{\dd F}{\dd L} .
\end{equation}
A number of the following applications will be discussed within the
zero temperature limit, with a change of notation from the free energy $\calF$
to the ordinary energy $\calE$ at zero temperature.

\section{Applications to non trivial geometries}

Formula (\ref{CasimirEnergyTnull}) has been used to evaluate the effect of roughness
or corrugation of the surfaces on the value of the Casimir force
\cite{Maia05,Rodrigues06,Rodrigues07,vanZwol08} in a
perturbative manner with respect to the roughness or corrugation amplitudes.
It has also allowed one to study a Bose-Einstein condensate used
as a local probe of vacuum above a nano-grooved plate \cite{Dalvit08,Messina09}.
The scattering approach has clearly a larger domain of applicability, not limited
to the perturbative regime, as soon as techniques are available for computing
the large matrices involved in its evaluation
\cite{Lambrecht08,Lambrecht08b,Chiu10}.

Another important application, which we will summarize also in the present section,
corresponds to the plane-sphere geometry used in most Casimir force experiments
and for which explicit ``exact calculations" (see a discussion of the meaning of this
expression below) have recently become available
\cite{Emig08,Maia08,CanaguierPRL09,CanaguierPRL10,CanaguierPRAsubmitted}.

\subsection{Perturbative treatment of shallow corrugations}

As already stated, the lateral Casimir force between corrugated plates
is a topic of particular interest. It could in particular allow for a new
test of Quantum ElectroDynamics through the dependence of
the lateral force on the corrugation wavevector
\cite{Rodrigues06,Rodrigues07}.

\begin{figure}[h]
\centering
\includegraphics[width=0.3\textwidth]{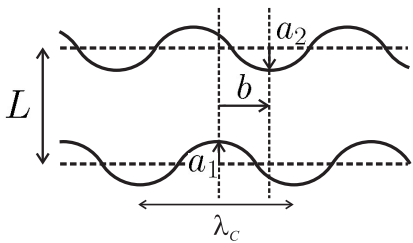}
\caption{Parallel corrugated surfaces, with $L$ representing the mean separation distance,
$a_1$ and $a_2$ the corrugation amplitudes and $b$ the lateral mismatch between
the crests. When the corrugation are supposed to be the smallest length
scales, the effect of the corrugations can be studied in the perturbative expansion.
This approximation will be dropped later on.}
\label{lateral-fig1}
\end{figure}

Here, we consider a geometry with two plane mirrors, $M_1$ and $M_2$, having corrugated surfaces
described by uniaxial sinusoidal profiles such as shown in Fig. \ref{lateral-fig1}.
We denote $h_1$ and $h_2$ the local heights with respect to mean planes $z_1=0$ and $z_2=L$
\begin{eqnarray}
&&h_1=a_1\,\cos(\kC x) ~ , ~ h_2=a_2\,\cos\left(\kC (x-b)\right) ~, \\
&&\kC=2\pi/\lambda_\C ~ . \nonumber
\end{eqnarray}
$h_1$ and $h_2$ have null spatial averages and $L$ is the mean distance between the two surfaces;
$h_1$ and $h_2$ are both counted as positive when they correspond to a decrease in the separation;
$\lambda_\C$ is the corrugation wavelength, $\kC$ the corresponding wavevector,
and $b$ the spatial mismatch between the corrugation crests.
At lowest order in the corrugation amplitudes, when
$a_1, a_2 \ll \lambda_\C, \lambda_\P, L$ (with $\lambda_\P$ the
plasma wavelength describing the properties of the metallic mirror),
the Casimir energy may be obtained by expanding up to second order
the general formula (\ref{CasimirEnergyTnull}). This perturbative approximation will
be dropped in the next subsection.

The part of the Casimir energy able to produce a lateral force
is then found to be
\begin{eqnarray}
F^\mathrm{lat} &=& -\frac{\partial \delta \calE^\mathrm{corrug} }{\partial b} , 
\label{Ecorrug}
\\
\delta \calE ^\mathrm{corrug}  &=&
- \hbar \int_0^{\infty} \frac{\dd\xi}{2\pi}
\Tr \left( \delta\calR_1 \frac{\exp^{-\calK L}}{\calD_\mathrm{plane}  }
\delta\calR_2 \frac{\exp^{-\calK L}}{\calD_\mathrm{plane}  }  \right) . \nonumber
\end{eqnarray}
$\delta\calR_1$ and $\delta\calR_2$ are the first-order variation
of the reflection matrices $\calR_1$ and $\calR_2$ induced by the
corrugations;
$\calD_\mathrm{plane}  $ is the matrix $\calD$ evaluated at zeroth order in the
corrugations; it is diagonal on the basis of plane waves and
commutes with $\calK$.

Explicit calculations of (\ref{Ecorrug}) have been performed for the simplest
case of experimental interest, with two corrugated metallic plates
described by the plasma dielectric function.
These calculations have led to the following expression of the lateral part
of the Casimir energy
\begin{eqnarray}
\delta E^\mathrm{corrug}  = \frac A2 \GC(\kC) a_1 a_2 \cos (\kC b) .
\label{spectral}
\end{eqnarray}
The spectral sensitivity function $\GC(\kC)$ has been given and discussed
in \cite{Rodrigues07}.
Using its expression, it is possible to prove a properly defined
``Proximity Force Theorem'' which states that the PFA is recovered
at the limit of long corrugation wavelengths $\kC\to0$.
Obviously, this theorem does not imply that the PFA is always valid or,
in other words that $\GC(\kC)$ may be replaced by $\GC(0)$.

\begin{figure}[h]
\centering \includegraphics[width=0.45\textwidth]{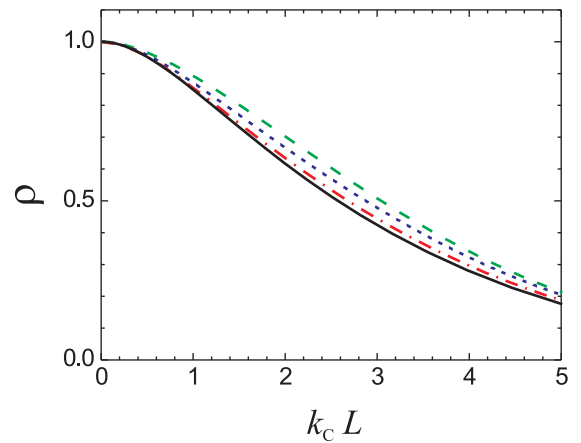}
\caption{Variation of $\rho$ versus the dimensionless variable $\kC L$
 for metallic mirrors described by the plasma model,
for $\kP L=$1 (dashed line), 2.5 (dotted line), 5 (dashed-dotted line)
and  10 (solid line) [colors online with respectively green, blue,
red and black lines]. }
\label{lateral-fig2}
\end{figure}

To assess the validity of the PFA for the lateral Casimir force description,
we now introduce the dimensionless quantity
\begin{eqnarray}
\rho(\kC)=\frac{\GC(\kC)}{\GC(0)} .
\label{rhocorrug}
\end{eqnarray}
The variation of this ratio $\rho$ with the parameters of interest
is shown in Fig. \ref{lateral-fig2}
for gold covered plates with $\lambda_\P=137$nm.
The ratio $\rho$ is smaller than unity as soon as $\kC$ significantly deviates from 0
which means that the PFA overestimates the lateral Casimir force.
For large values of $\kC$, it even decays exponentially to zero,
leading to an extreme deviation of the real lateral force from the PFA prediction.

Another situation of interest arises when the corrugation plates are rotated with respect to each other.
Assuming as previously corrugations of sinusoidal shape with corrugation wavevectors $\mathbf{k}_j$ having
the same modulus $k=2\pi/\lambda_C$ on both plates, it is possible to derive the second-order correction
$\delta E^\mathrm{torque}$ to the Casimir energy which depends on the angle $\theta$ between the
corrugations and thus has the ability to induce a Casimir torque \cite{Rodrigues06,RodriguesEPL06}.
Only crossed terms, proportional to the corrugation amplitudes on both plates, contribute to this expression,
as the square terms are independent of the angle $\theta$.
The expression $\delta E^\mathrm{torque}$ contains as the special case $\theta=0$
the pure lateral energy discussed above.
Note that the dependence on the material properties and corrugation wavevector
are captured by the same response function $\GC$ already calculated.

\begin{figure}[h]
\centering{\includegraphics[width=6cm]{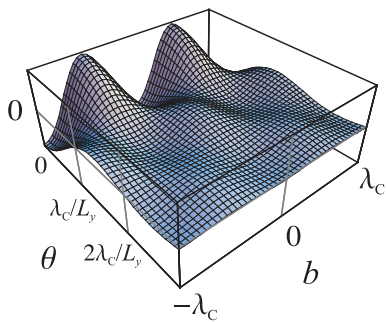}}
\caption{ Casimir energy (arbitrary unit) as a function of the rotation
angle $\protect\theta$ and the lateral displacement $b.$}
\label{torque-fig1}
\end{figure}

For quantitative estimations, we assume
that the corrugations are restricted to a rectangular section of area $L_x L_y$ with
transverse dimensions $L_x$ and $L_y$ much larger than the plate separation $L$ and neglect
diffraction at the borders of the plates.
In Fig.~\ref{torque-fig1}, we plot $\delta E^{\mathrm{torque}}$ obtained in this manner,
in arbitrary units, as a function of $b$ and $\theta.$
The Casimir energy is found to be minimal at $\theta=0$ and
$b=0,\lambda_C,2\lambda_C,...,$, which corresponds to a situation where corrugations are aligned.
Starting from $\theta=b=0$ and rotating
plate 2 around its center, one follows the line $b=0$ in Fig.~\ref{torque-fig1}.
Clearly, for small angles the plate is attracted back to $\theta=b=0$ without sliding laterally.

The Casimir torque is then deduced by deriving the energy with respect to the angle $\theta$
\begin{equation}
\tau =-\frac{\partial }{\partial \theta }\, \delta E^\mathrm{torque} .
\end{equation}
It is maximum at $\theta= 0.66\lambda_C/L_y$ where it is given by
\begin{equation}  \label{torque}
\frac{\tau}{L_xL_y}= 0.109\, a_1a_2\, k {\GC}(k)\,L_y.
\end{equation}
The maximum torque per unit area is proportional to
the length $L_y$ of the corrugation lines, which plays the role of the
moment arm.

In contrast with the similar torque appearing between misaligned
birefringent plates \cite{Munday05}, the torque is here coupled to
the lateral force. This could induce complicated behaviours in an experiment
and would probably have to be controlled.
This can be clearly seen on Fig. \ref{torque-fig1}: if the plate is
released after a rotation of $\theta > \lambda_C/L_y$ it will move
in a combination of rotation and lateral displacement. The
energy correction vanishes at $\theta=\lambda_C/L_y$, defining the range of
stability of the configuration $b=\theta=0.$ Rotation is favored over
lateral displacements only for $\theta <\lambda_C/L_y.$

\begin{figure}[h]
\centering{\includegraphics[width=8cm]{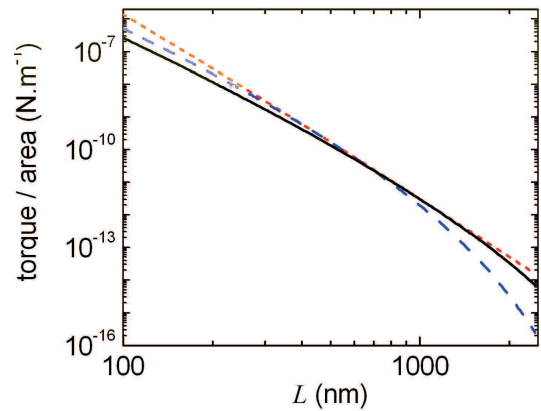}}
\caption{Maximum torque per unit area as a function of the mean separation $L$ for the following
parameters: $a_1a_2=200\,\mathrm{nm}^2,$ $L_y=24\,\protect\mu\mathrm{m},$ $\protect\lambda_P=137\,%
\mathrm{nm}.$ Solid line: $\protect\lambda_C=2.4\,\protect\mu\mathrm{m};$
dashed line: $\protect\lambda_C=1.2\,\protect\mu\mathrm{m};$ dotted line: $%
\protect\lambda_C=2\protect\pi L/2.6$ (corresponding to the optimum value). }
\label{torque-fig2}
\end{figure}

However, the advantage of the configuration with
corrugated plates is that the torque has a larger magnitude.
Fig. \ref{torque-fig2} shows the maximum torque as a function of mean separation
between the two corrugated gold plates with  a plasma wavelength $\lambda_P=137$nm.
At a plate separation of about 100nm the torque per unit area can be as high as $10^{-7}$Nm$^{-1}$
These results on lateral forces and Casimir torques suggest that non trivial effects of geometry,
\textit{i.e.} effects beyond the PFA, can be observed with dedicated experiments.
It is however difficult to achieve this goal with corrugation amplitudes $a_1, a_2$
meeting the conditions of validity of the perturbative expansion.
This approximation is dropped in the next subsection.

\subsection{Non perturbative calculations with deep gratings}

As already stated, recent experiments have been able to probe the beyond-PFA regime
with deep corrugations \cite{Chan08,Chiu09} and it has also become possible to calculate
exact expressions of the forces between nanostructures without using the perturbative
assumption. This necessarily involves the non specular scattering formula
(\ref{CasimirFreeEnergyNS}) and the evaluation of scattering properties mixing
different wavevectors and polarizations.

\begin{figure}[h]
\centering
\includegraphics[width=4cm]{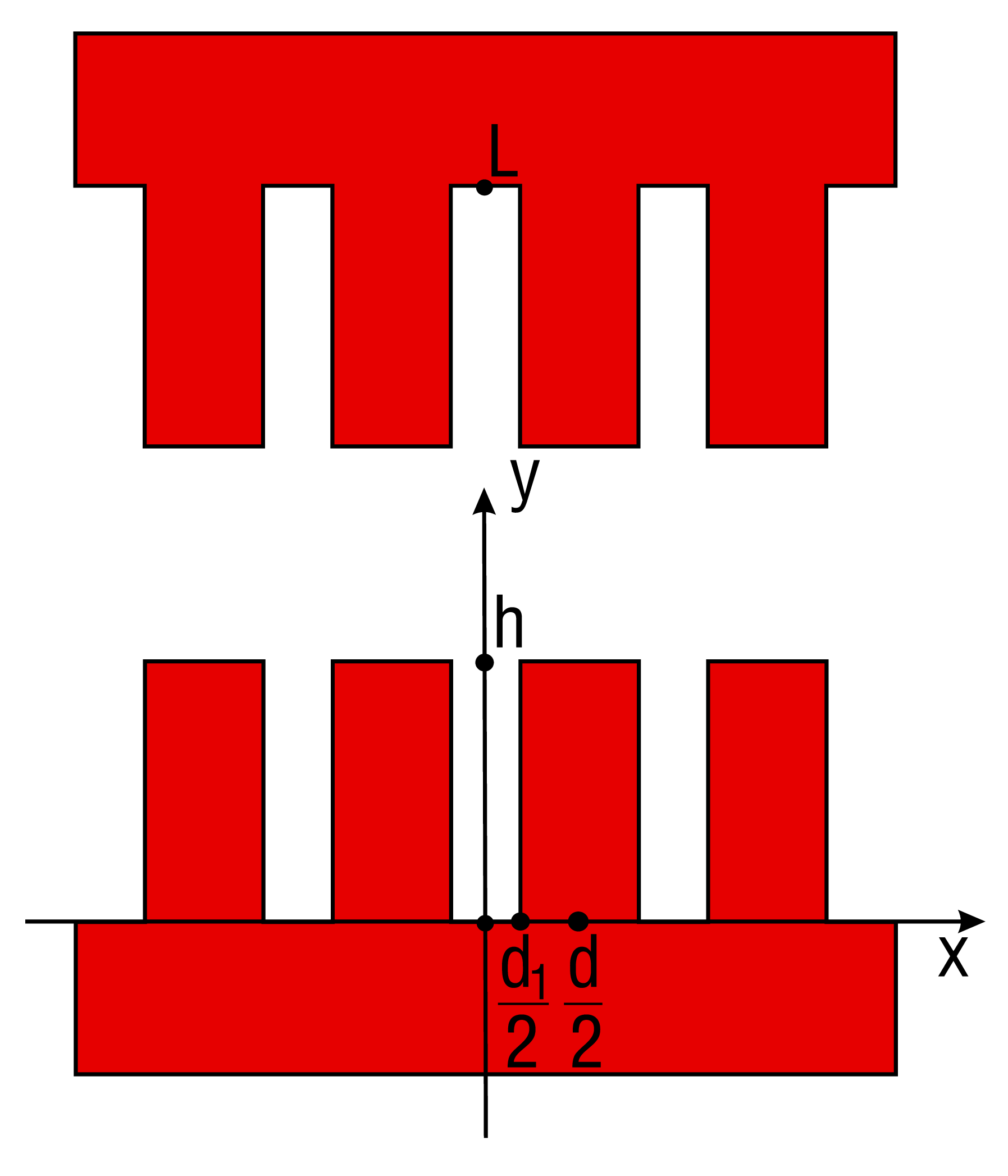}
\caption{Two surfaces with rectangluar gratings of depth $h$, gap width $d$ and trench width $d-d_1$.}
\label{gratings-fig1}
\end{figure}

In the following we briefly discuss the Casimir interaction energy in a typical
device made of two nanostructured surfaces of intrinsic Silicon, such as shown in Fig. \ref{gratings-fig1}.

To model the material properties of intrinsic Silicon, we use a Drude-Lorentz model
for which the dielectric function is well approximated by \cite{Bergstrom97}
\begin{eqnarray}
\varepsilon(i\xi) = \varepsilon_{\infty}+
\frac{(\varepsilon_0-\varepsilon_{\infty}) \xi_0^2}{\xi^2 +
\xi_0^2}\label{epsilonapprox} ,
\end{eqnarray}
with $\varepsilon_0 \approx 11.87$ the value of the dielectric function at zero frequency,
$\varepsilon_{\infty} \approx 1.035$ the high frequency limit of the dielectric function
and $\omega_0 = i\xi_0 \approx 4.34$ eV. Calculated with the proximity force approximation,
the Casimir force between the two gratings is given by the geometric sum
of two contributions corresponding to the Casimir force between two
plates $F_{PP}$ at distances $L$ and $L-2h$, which is independent of the corrugation period $d$.

\begin{figure}[h]
\centering
\includegraphics[width=6cm]{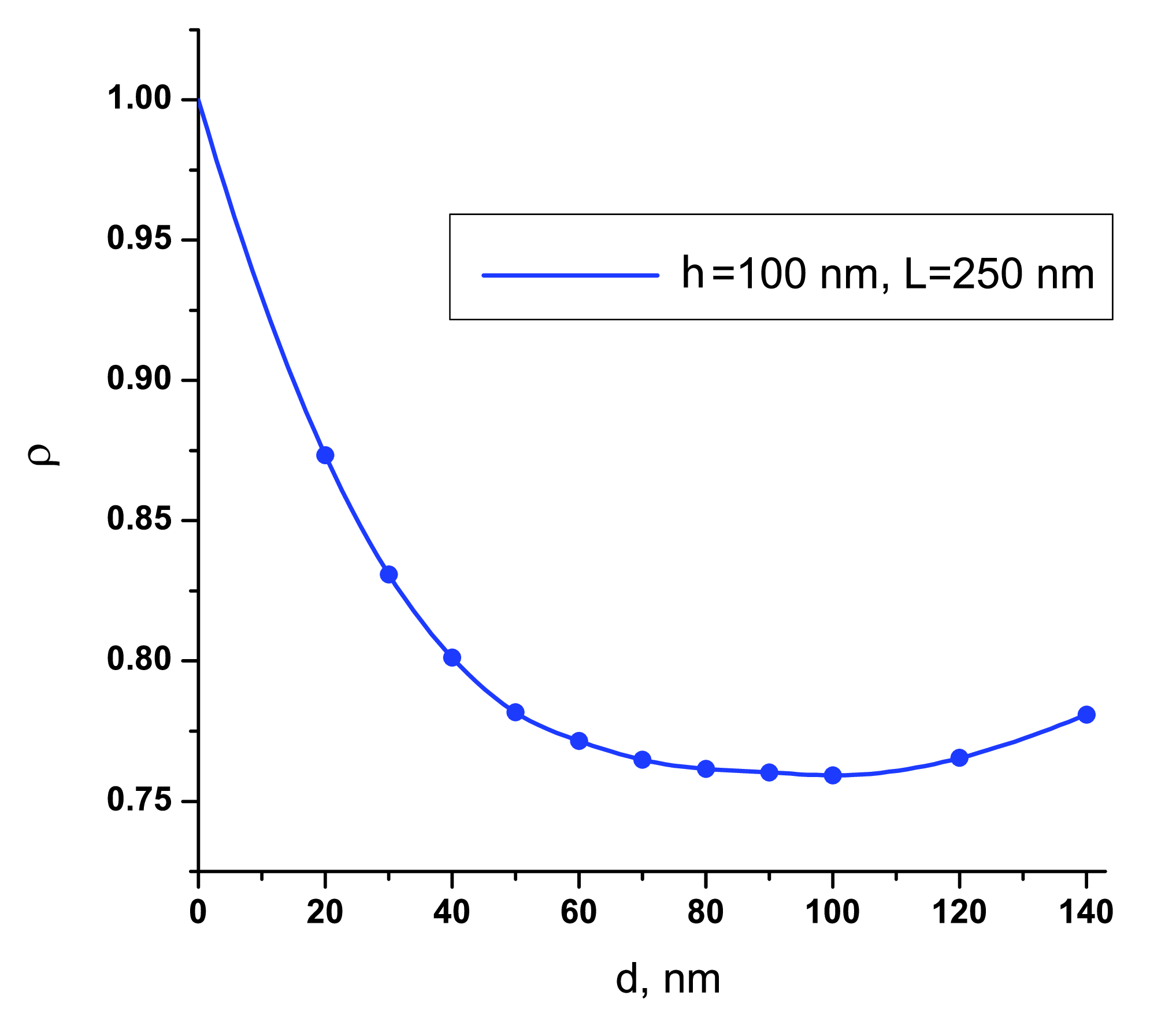}
\caption{Casimir force normalized by its PFA value for two
gratings of intrinsic Silicon with amplitude $h=100$nm and $d_1=\frac{d}{2}$
as a function of $d$
at a fixed distance $L=250$nm.}
\label{gratings-fig2}
\end{figure}

To assess quantitatively the validity of PFA, we plot as before the dimensionless
quantity
\begin{equation}
\rho=\frac{F}{F_\textrm{PFA}} .
\end{equation}
Fig. \ref{gratings-fig2} shows this ratio for two Silicon gratings,
separated by $L=250$nm, of height $h=100$nm as a function of the corrugation
period $d$ with $d_1=d/2$ \cite{Lambrecht08}.
Clearly, PFA is not a valid approximation except for two limiting cases,
that is a vanishing corrugation period $d \rightarrow 0$ and a very large
corrugation periods $d \rightarrow \infty$, meaning in either case that
the structured surfaces becomes flat. In between the exact result for the
Casimir force is always smaller than the PFA prediction, meaning that PFA
overestimates the force. This has to be contrasted with calculations for
perfect conductors where PFA always underestimates the real force.

\begin{figure}[htbp]
  \begin{center}
    \includegraphics[width=8cm]{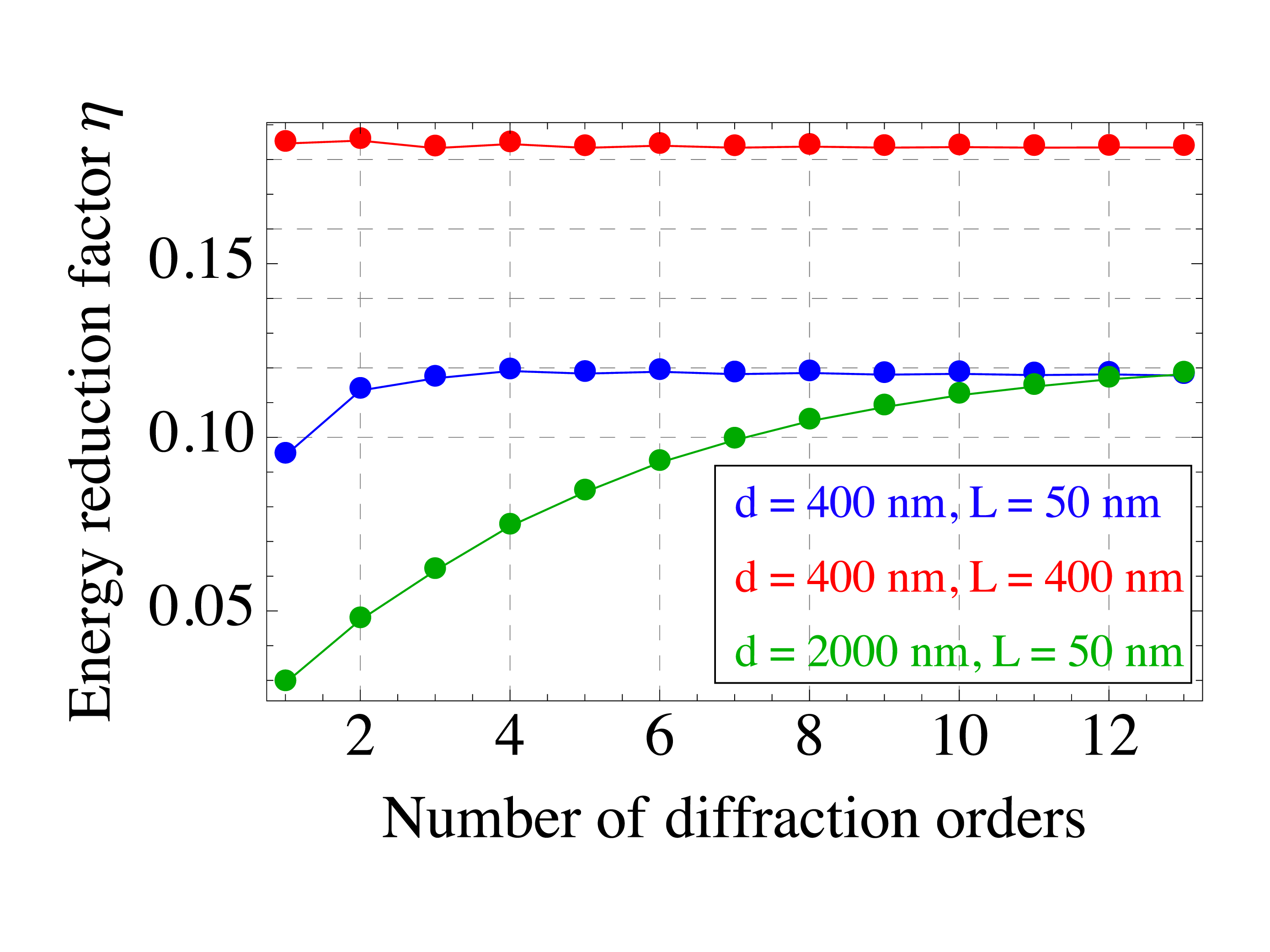}
  \end{center}
  \caption{Convergence of the calculated Casimir energy between two gratings as a function of the
  number of diffraction orders retained in the calculation.
  Gratings with different periods are plotted as blue and red ($400$nm) and green ($2\ \mu$m) points.
  The convergence of the calculations becomes slower when increasing the grating period $d$ or decreasing the separation $L$.}
  \label{fig:convN}
\end{figure}

One important parameter to keep in mind is the number of diffraction orders that has to be retained in the calculation
in order for the Casimir energy to converge in the numerical calculation. This is illustrated in Fig.~\ref{fig:convN} for two
Silicon gratings. For the sake of convenience, we plot the Casimir energy normalized by the energy for perfectly reflecting plane
mirrors \textit{i.e.} the energy reduction factor.
The blue curve corresponds to the situation of two gratings of period $400$nm separated by
a distance $L=50$nm. Clearly around five orders of diffraction are sufficient for the calculation of the Casimir energy in this case. The
number of necessary diffraction orders decreases with increasing distance between the gratings. This is illustrated
by the red curve where the two aforementioned gratings are now separated by a distance $L=400$nm and where the
Casimir energy has basically converged to its final value with only one order of diffraction retained. The fast convergence is here due to that fact that oblique diffraction orders are exponentially suppressed with increasing distance \cite{Messina09}.
Finally, the greater the period of the grating the more orders of diffractions are needed as shows the green curve where the period of the two gratings is now $2\mu$m. In this case, the Casimir energy has not yet fully converged to its final value even with as much as 13 orders of diffraction. This can be understood because the momentum transferred by the grating $q=\frac{2 \pi}{d}$ is now small so that different orders of diffraction are nearly collinear with the specular one and therefore greatly contribute to the final energy.

\begin{figure}[h]
  \begin{center}
    \includegraphics[width=8cm]{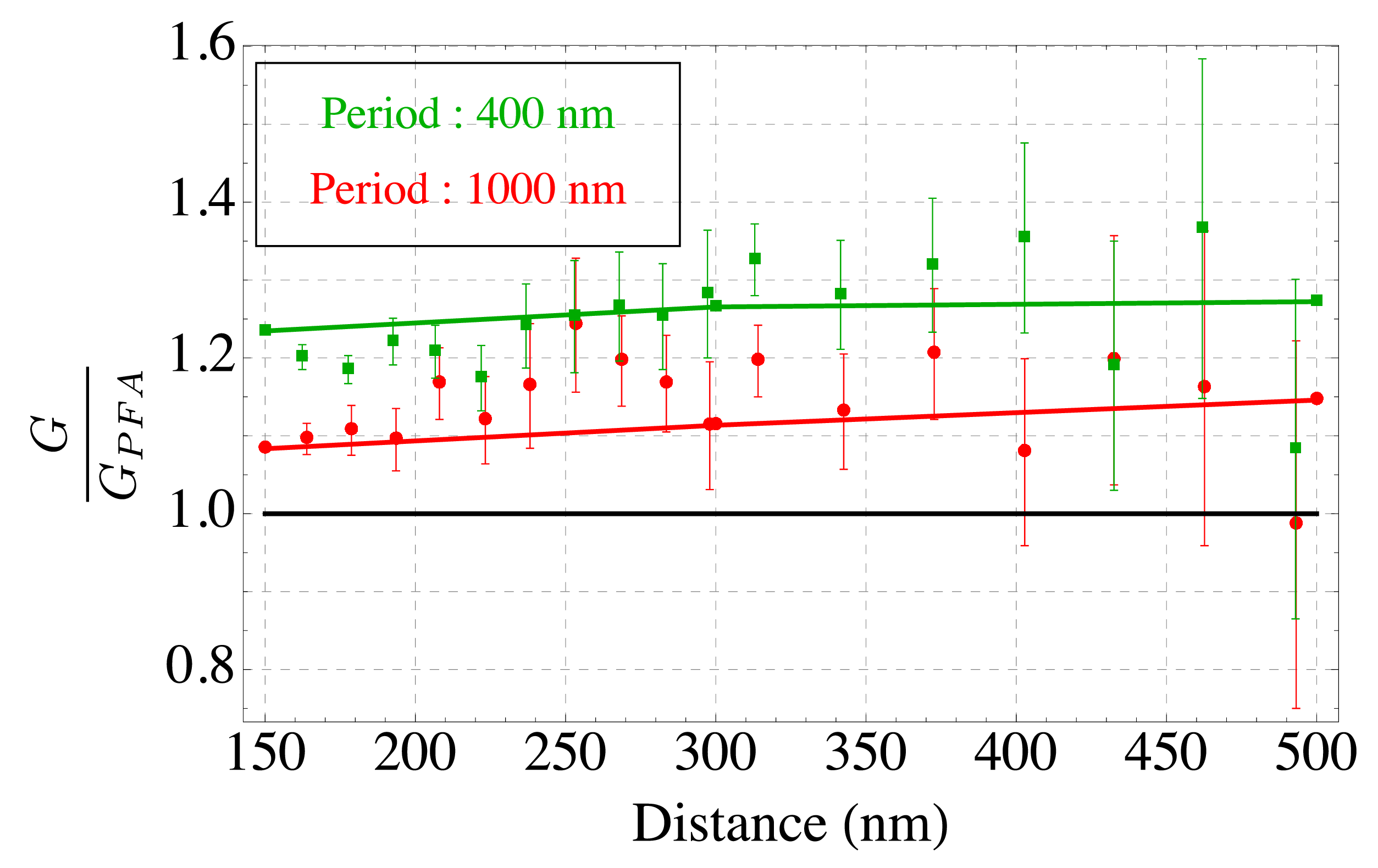}
  \end{center}
  \caption{Comparisons between experimental measurements and exact calculations for the Casimir force
  gradient between a gold sphere and two types of silicon gratings. Green and red dots correspond to data points provided by Ho Bun Chan for a grating period of $400$nm and $1\mu$m respectively. The solid curves of the same color are calculated data obtained using the scattering approach for the corresponding experimental parameters.}
  \label{fig:CompExp}
\end{figure}

If attention is paid to the issue of convergence this calculation method is essentially exact and allows for direct comparisons with experimental
results. In a recent experiment, Chan et al. have measured the Casimir force gradient between a gold sphere and a grating of doped silicon \cite{Chan08}. Two samples of silicon gratings have been used. Both have a corrugation
depth of $1\mu$m, but different periods of $400$nm and $1\mu$m respectively. The experimental data points of the ratio between the force gradient and its PFA approximation for both samples have been kindly provided by Ho Bun Chan and are plotted in Fig.~\ref{fig:CompExp}.

Concerning the calculation we model the optical properties of Silicon by the dielectric function (\ref{epsilonapprox}). We have also taken into account the doping of the Silicon by adding a Drude part to this dielectric function, but this has not led to noticeable changes for the Casimir interaction in the distance range up to 500nm which has been explored in the experiment. To model the optical properties of gold we have used available optical data, extrapolated at low frequencies by a Drude model
$\epsilon(i \xi)=\frac{\omega_{p}^{2}}{\xi(\xi+\gamma)}$ with $\omega_{p}=9$eV and
$\gamma=35$meV. The method is described in detail in \cite{LambrechtEPJ00}. The calculations were run up to $N=3$ diffraction orders, after which the result for the Casimir energy was found to have converged. The result of our calculation is given in Fig.~\ref{fig:CompExp} as the solid green and red curves for the $400$nm and $1\mu$m samples respectively. The theoretical predictions and the experimental data points are in good agreement. Due to a new improved numerical code the agreement is better than the one presented in \cite{Lambrecht08}.

\subsection{Exact calculations in the plane-sphere geometry }

The plane-sphere geometry is the configuration in which the most precise
Casimir force measurements are currently performed \cite{Krause07}.
The Casimir interaction in this geometry can also be calculated in a formally exact
manner using the general scattering formula (\ref{CasimirFreeEnergyNS}).
Such calculations have first been performed for perfectly reflecting mirrors
\cite{Emig08,Maia08} where it was found that the Casimir energy was smaller
than expected from the PFA and, furthermore, that the result for
electromagnetic fields was departing from PFA more rapidly than was expected
from previously existing scalar calculations \cite{BordagJPA08,WirzbaJPA08}.
It is only very recently that the same calculations have been performed for the
more realistic case of metallic mirrors at zero temperature \cite{CanaguierPRL09}
and at arbitrary temperature \cite{CanaguierPRL10,CanaguierPRAsubmitted} where both the lossless plasma model
dielectric function and the lossy Drude dielectric function have been studied.
We will sketch the method in the following.

\begin{figure}[h]
\centering
\includegraphics[width=4cm]{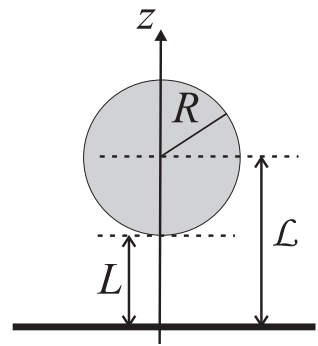}
\caption{The geometry of a sphere of radius $R$ and a plate
at distance $L$; the center-to-plate distance is $\cL\equiv L+R$.}
\label{Plane-sphere}
\end{figure}

The set-up of a sphere of radius $R$ above a flat plate
is schematically presented in Fig.~\ref{Plane-sphere}.
We denote respectively $L$ and $\cL\equiv L+R$ the closest approach distance
and the center-to-plate distance.
In this configuration, the general expression of the Casimir free energy
at temperature $T$ may be written as
\begin{eqnarray}
\label{depart}
&&\calF = k_\B T \sum_m{}^\prime \Tr \ln \calD (i\xi_m) 
\quad,\quad \\
&& \calD \equiv 1-\cR_\S e^{-\cK \cL} \cR_\P e^{-\cK \cL} . \nonumber
\end{eqnarray}
The expression contains the reflection operators of the sphere $\cR_\S$ and the
plate $\cR_\P$ which are evaluated with reference points placed at the sphere
center and at its projection on the plane of the plate.
They are sandwiched in between operators $e^{-\cK \cL}$ describing the propagation
between the two reference points.

The upper expression is conveniently written through a decomposition
on suitable plane-wave and multipole basis \cite{CanaguierPRL09};
$\cR_\P$ is thus expressed in terms of the Fresnel reflection coefficients
$r_p$ with $p=$TE and TM for the two electromagnetic polarizations, while
$\cR_\S$ contains the Mie coefficients $a_\ell,b_\ell$
for respectively electric and magnetic multipoles at order $\ell=1,2,...$.
Due to rotational symmetry around the $z$-axis, each eigenvalue of the
angular momentum $m$ gives a separate contribution to the Casimir free
energy $\cF^{(m)}$, obtained through the same formula as (\ref{depart}).
The scattering formula is obtained by writing also transformation formulas
from the plane waves basis to the spherical waves basis and conversely.

The result takes the form of a multipolar expansion with spherical waves
labeled by $\ell$ and $m$ ($\vert m\vert\le\ell$). It can be considered
as an ``exact'' multipolar series of the Casimir free energy.
Of course, the numerical computations of this series can only be done
after truncating the vector space at some maximum value $\ell_\max$
of the orbital index $\ell$.

\begin{figure}[h]
\centering
\includegraphics[width=7cm]{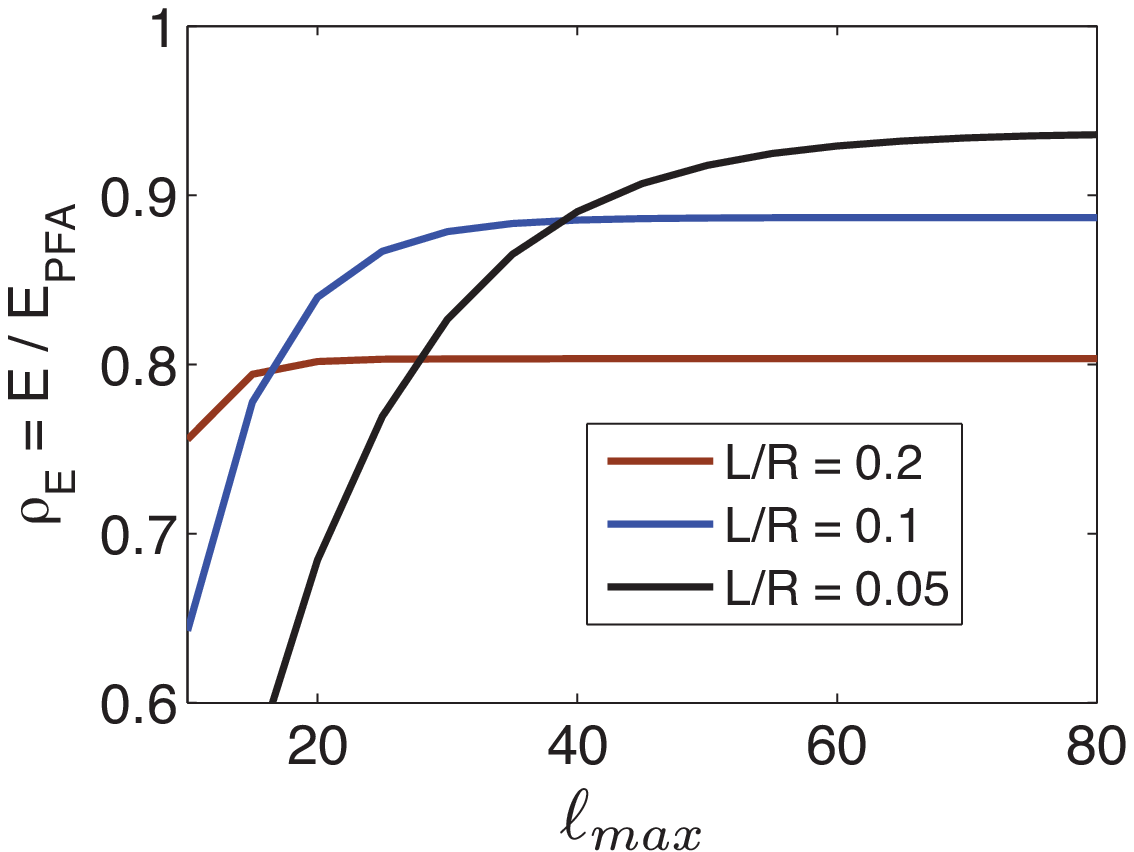}
\includegraphics[width=7cm]{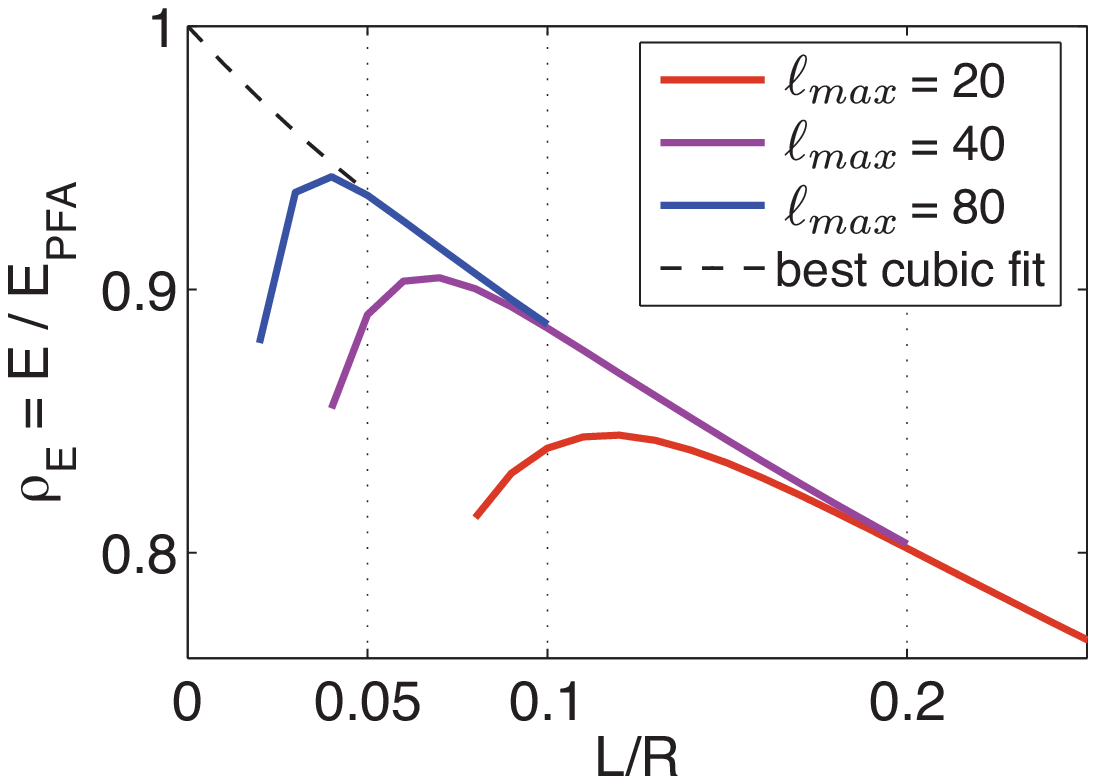}
\caption{Upper graph:  the ratio $\rho_E={E}/{E^\PFA}$ of the plane-sphere Casimir energy
to its PFA estimation is plotted as a function of $\ell_\max$ for different values of $L/R=0.05, 0.1, 0.2$.
Lower graph: same ratio $\rho_E$ plotted as function of $L/R$ for different values of $\ell_\max=20,40,80$.}
\label{EffectEllMax}
\end{figure}

The effect of this truncation is represented on Fig. \ref{EffectEllMax}
where the Casimir energy in the plane-sphere geometry divided by its PFA estimation
\begin{eqnarray}
\rho_E=\frac{E}{E^\PFA}
\end{eqnarray}
is plotted for various values of $\ell_\max$, in the special case of
perfect mirrors in vacuum ($T=0$).
The figure shows that as expected the numerical results are more and more
accurate when $\ell_\max$ is increased.
More precisely the accuracy is significantly degraded when the ratio $L/R$
goes below a minimal value inversely proportional to $\ell_\max$
\begin{eqnarray}
x\equiv \frac {L}{R} > x _{\min} \quad,\quad
x_{\min}\propto \frac {1}{\ell_\max} .
\end{eqnarray}
To illustrate the effect of the truncation, one may say that the accuracy
is degraded by typically more than $0.1\%$ when $x<0.05$ for a value of $\ell_\max=$85.
For small values of $x$, which corresponds to the most precise current
experiments, it may be possible to obtain information through
an extrapolation of the numerical results. As an example, the dashed line
on Fig. \ref{EffectEllMax} shows the result of a third degree polynomial
fit using accurate numerical evaluations.

\begin{figure}[h]
\centering
\includegraphics[width=8cm]{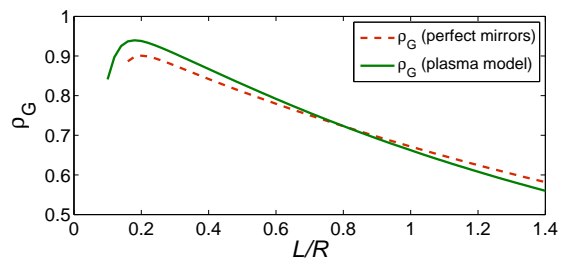}
\caption{Variation of
$\rho_\G$ as a function of $L/R$ as a function of
$L/R$, for a nanosphere of radius $R=100$nm; the solid green line
corresponds to gold-covered plates ($\lambda_\P=136$nm) and the
dashed red line to perfect reflectors.  The decrease at low values of $L/R$ represent a
numerical inaccuracy due to the limited value of $\ell_\max$ (24 in this calculation \cite{CanaguierPRL09}).}
\label{plane-sphere-fig2}
\end{figure}

As a further step, we show now on Fig. \ref{plane-sphere-fig2}
the results corresponding to perfect and plasma mirrors,
still at zero temperature \cite{CanaguierPRL09}.
We have derived the Casimir energy (\ref{depart}) to obtain expressions
for the force $F$ and force gradient $G$, and then formed the ratios
of the plane-sphere exact results to the PFA expectations $F^\PFA$
and $G^\PFA$ respectively
\begin{eqnarray}
\rho_F=\frac{F}{F^\PFA} \quad,\quad \rho_G=\frac{G}{G^\PFA} .
\end{eqnarray}

Using these theoretical evaluations, it is now possible to extract
information of interest for a comparison with
the experimental study of PFA in the plane-sphere geometry \cite{Krause07}.
In this experiment, the force gradient has been measured for various radii of the
sphere and no deviation of PFA was observed.
The authors expressed their result as a constraint
on the slope at origin $\beta_G$ of the function $\rho_G(x)$
\begin{equation}
\rho_G(x)=1+\beta_G x+O(x^2) 
\quad,\quad
\vert\beta_G\vert<0.4 .
\end{equation}
Reasoning along the same lines, we have interpolated our theoretical evaluation of
 $\rho_G$ at low values of $x=L/R$ \cite{CanaguierPRL09}.
Surprisingly the slope obtained for perfect reflectors was found to
lie outside the experimental bound of \cite{Krause07}
\begin{equation}
\beta_G^\perf \sim-0.48 .
\end{equation}
The consistency with this bound is however recovered for the
calculations done for plasma mirrors
\begin{equation}
\beta_G^\plas \sim-0.21 .
\end{equation}

\begin{figure}[h]
\centering
\includegraphics[width=8cm]{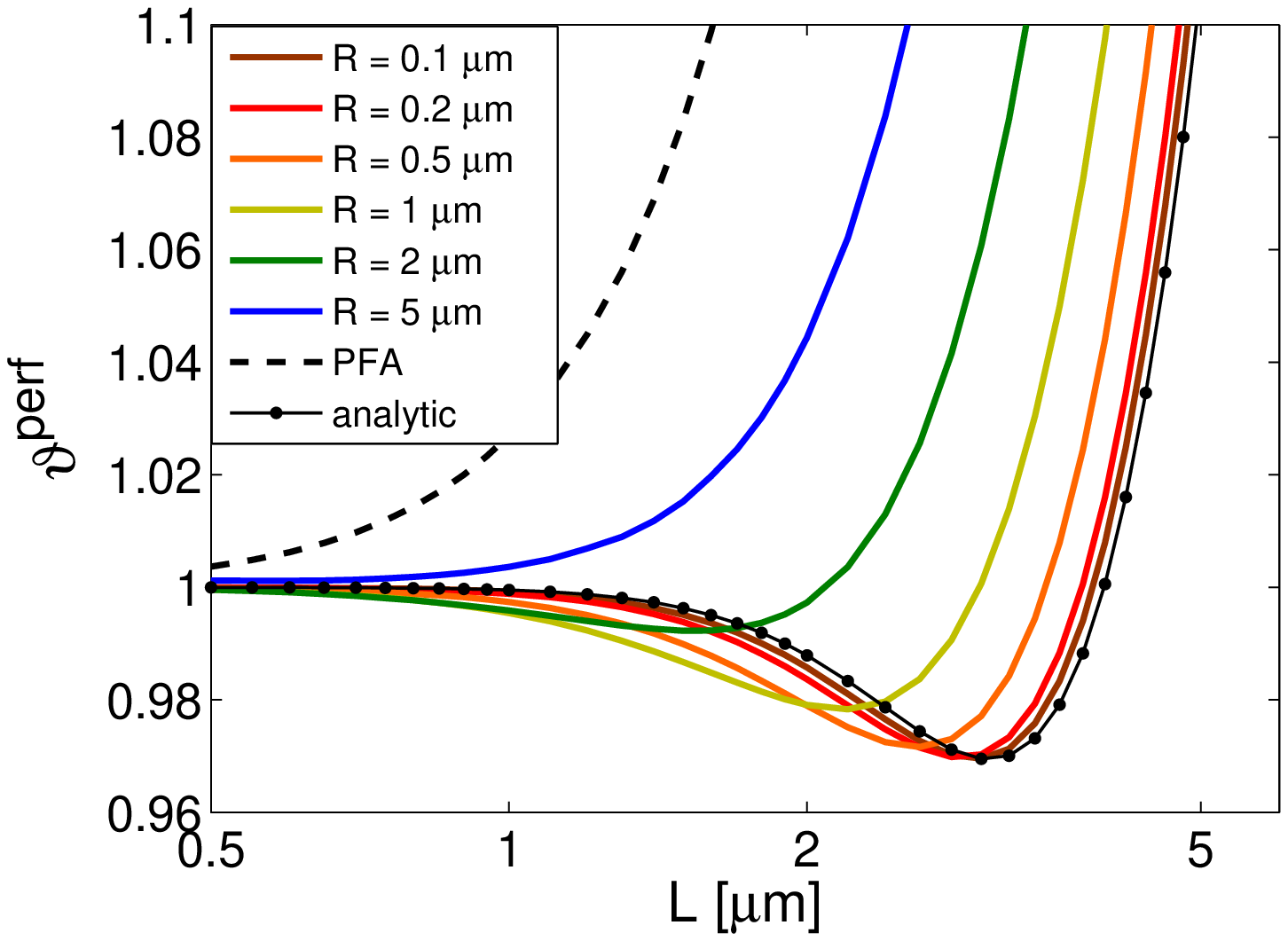}
\includegraphics[width=8cm]{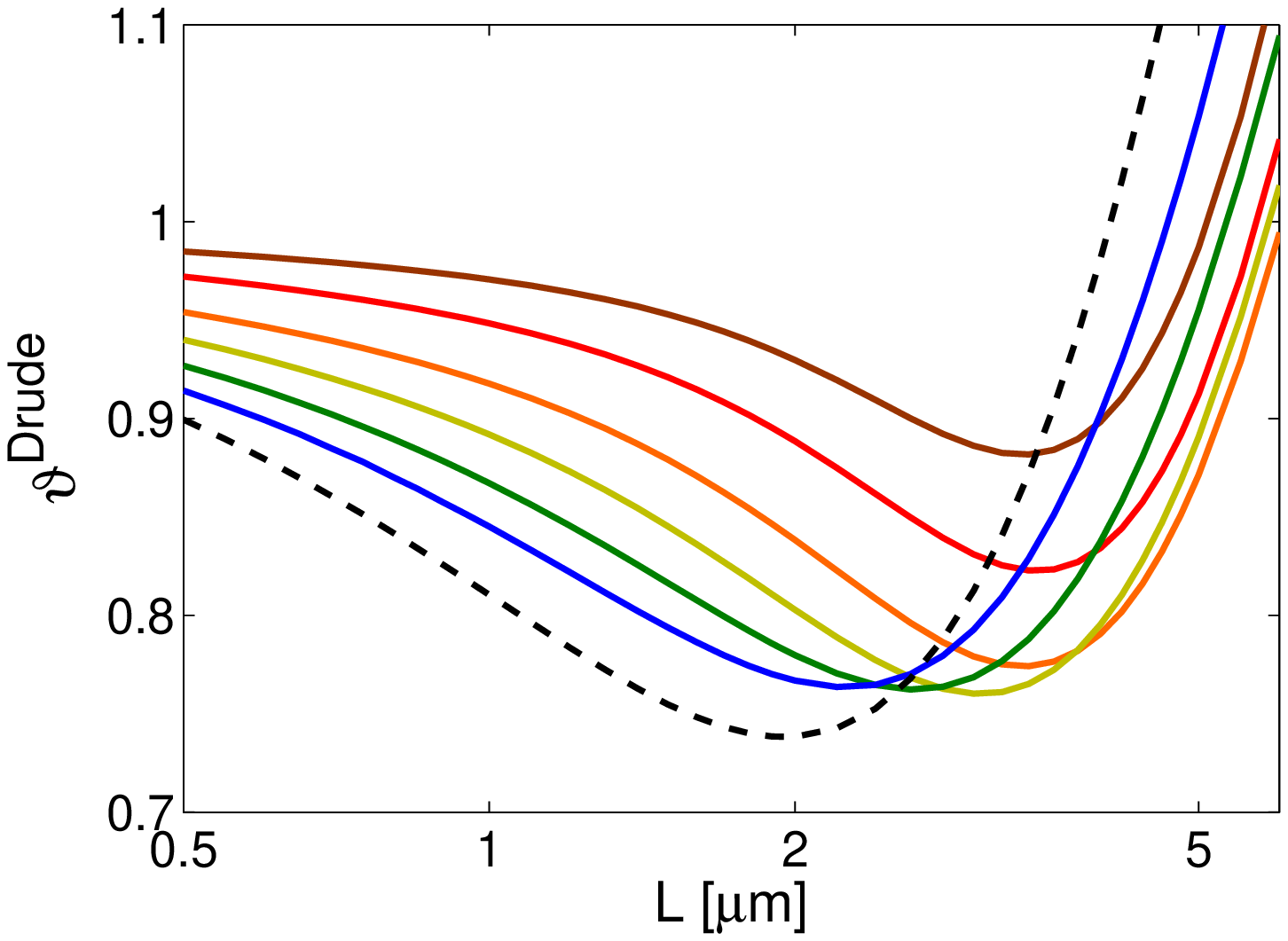}
\caption{Thermal Casimir force at $T=300$K divided by the zero temperature force,
computed between perfectly reflecting
sphere and plane (upper graph), and between Drude metals (lower graph) plotted for
$\lambda_\P=136$nm, $\lambda_{\gamma}/\lambda_\P=250$. The solid lines from bottom
to top correspond to increasing values of sphere radii.
The dotted curve in the upper graph is the analytical asymptotic expression in the $L \gg R$ limit.
The PFA expressions are given by the dashed curves.} 
\label{plane-sphere-T-fig1}
\end{figure}

As a last example of application,
 we now discuss the effect of a non zero temperature.
To this aim we evaluate eqn (\ref{depart}) at ambient temperature ($T=300$K).
The results of the numerical computations are shown on Fig. \ref{plane-sphere-T-fig1},
for the limiting case of perfect reflection (left) and for Drude metals (right)
evaluated for $\lambda_\P=136$nm, $\lambda_{\gamma}/\lambda_\P=250$
(values corresponding to gold).
We have calculated the Casimir force $F^\perf$ and $F^\Drud$
between the plane and the sphere at ambient temperature and
then plotted the corresponding ratios $\vartheta^\perf$ and $\vartheta^\Drud$ of
this force to a reference force
corresponding to zero temperature
\begin{eqnarray}
\label{deftheta}
F^\perf(L,T)&\equiv&-\frac{\partial\cF^\perf}{\partial L} ~ , ~
\vartheta^\perf\equiv\frac{F^\perf(L,T)}{F^\perf(L,0)} ,   \\
F^\Drud(L,T)&\equiv&-\frac{\partial\cF^\Drud}{\partial L} ~, ~ 
\vartheta^\Drud\equiv\frac{F^\Drud(L,T)}{F^\Drud(L,0)} .\nonumber
\end{eqnarray}
The various solid curves are drawn for different
sphere radii $R$ as a function of the separation $L$~;
the dashed curves on Fig.~\ref{plane-sphere-T-fig1}
represent the quantities $\vartheta^{\perf}_{\PFA}$ and $\vartheta^{\Drud}_{\PFA}$
obtained from (\ref{deftheta}) by using the PFA~;
the dotted curve in the upper graph is an analytical asymptotic expression
discussed below. We do not show the corresponding plots for plasma mirrors as they are
very similar to the perfect mirror case.

The comparison of $\vartheta^{\perf}$ and $\vartheta^{\Drud}$ reveals surprising features,
which could not be expected from an analysis in the parallel-plate geometry.
First both ratios $\vartheta$ start from unity at small distances $L/R \rightarrow 0$.
For $R$ small enough, the ratios then decrease below unity with increasing distance, reach a radius-dependent
minimum and then increase again at very large distances. This behavior entails that the absolute value of the
Casimir force is smaller at $T=300$K than at $T=0$, implying a repulsive contribution
of thermal fluctuations. The dashed PFA curve in the
upper graph of Fig. \ref{plane-sphere-T-fig1}
representing $\vartheta^{\perf}_{\PFA}$ is always larger than unity, excluding
such a repulsive contribution of thermal fluctuations in the plane-plane geometry.

A second important feature showing up in Fig. \ref{plane-sphere-T-fig1} is that the PFA
expression always
overestimates the effect of temperature on the force between perfect
(and plasma) mirrors. However between Drude metals, the PFA underestimates this effect
at small distances and overestimates it at
large distances, the overestimation being however smaller than for
perfect mirrors. These results clearly indicate that there is a
strong correlation between the effects of plane-sphere geometry,
temperature and dissipation.

The calculation of the Casimir free energy may be
done analytically for small frequencies corresponding to large plane sphere separations
\begin{eqnarray}
\label{Eperfectanalytical}
&&\cF_\lone^\perf =
-\frac{3 \hbar c R^3}{4 \lambda_\T L^3} \phi(\nu) \quad,\quad
\nu \equiv \frac{2\pi L}{\lambda_\T} , \\
&&\phi(\nu) \equiv \frac{ \nu^2\cosh\nu + \nu\sinh\nu + \cosh\nu\sinh^2\nu}
{2 \sinh^3\nu} . \nonumber
\end{eqnarray}
This simple expression is a good approximation, as proven
by the fact that the full expression of $\vartheta^\perf$ tends indeed
asymptotically to this simple form for small radii $R \ll L$
(dotted line on upper graph of Fig. \ref{plane-sphere-T-fig1}).
One can also derive from this expression interesting information
about the behaviour of the Casimir entropy
\begin{eqnarray}
\label{Sperfectanalytical}
&&S_\lone^\perf = -\frac{\partial\cF_\lone^\perf }{\partial T}
=\frac{3 k_\mathrm{B} R^3}{4 L^4}
\left( \phi(\nu) + \nu \phi^\prime(\nu) \right)
\end{eqnarray}
This expression takes on negative values for $\nu\lesssim 1.5$, that
is $L\lesssim 1.8\mu$m at $T=300$K, which is in agreement with the behavior
observed in the upper graph of Fig. \ref{plane-sphere-T-fig1}: in most cases
$\vartheta^\perf$ decreases below unity as the distance
increases, reaches a minimum and then increases again at long distances.
As long as $R$ is not too large, the thermal photons provide a repulsive contribution
over a distance range that gets wider as $R$ decreases, to become $L \lesssim \lambda_T/2$ for very small spheres.

We finally will compare the predictions of the dissipation-less plasma model and
the dissipative Drude model for the thermal Casimir interaction in the plane-sphere geometry.
The difference will become particularly clear in the high temperature limit
$\cL \gg \lambda_T$ where one only needs to take the first Matsubara frequency $\xi_0=0$
when computing the Casimir free energy.
In the low frequency limit, the Fresnel coefficients (\ref{Fresnel}) for the plates are  given by
$r_{\rm TE}\approx - r_{\rm TM}\approx -1$ for the plasma model. The Mie coefficients are easily evaluated
\cite{CanaguierPRL10,CanaguierPRAsubmitted} and the following approximation for the Casimir force within the
plasma model
\begin{eqnarray}
\label{Mieplasma2}
&&\cF^\plas  \approx   - \frac{3 \hbar c R^3}{8 \lambda_T \cL^3}
\left( 1 + \frac{1}{\alpha^2}-\frac{\coth\alpha}{\alpha} \right) ~ ,~ \nonumber \\
Ê&&\cL \gg  \lambda_T, R , \lambda_P ~,   
\quad \alpha \equiv  \frac{2\pi R}{\lambda_P}. \nonumber
\end{eqnarray}
This result reproduces, as a particular case, the perfectly-reflecting limit when $\lambda_P\ll R.$

For the Drude model, the TE Fresnel reflection coefficient has the well-known
low-frequency limit $r_{\rm TE}\rightarrow 0,$ whereas
the TM coefficient behaves as in the plasma model:  $r_{\rm TM}\approx 1.$
The low-frequency expansion of the Mie coefficients are also quite different
from the plasma case and can be found in \cite{CanaguierPRL10,CanaguierPRAsubmitted}.
The resulting high-temperature large-distance limit for the free energy is
\begin{equation}\label{F_Drud}
\cF^\Drud \approx  - \frac{\hbar c R^3}{4 \lambda_T \cL^3} \quad , \quad  \cL \gg  \lambda_T, R .
\end{equation}
This remarkable result does not depend on the length scales $\lambda_P$ and
$\lambda_{\gamma}$ characterizing the material response, whereas
the corresponding plasma result (\ref{Mieplasma2}) clearly depends on $\lambda_P.$
One can show that
this is always the case in the high-temperature limit
$\lambda_T\ll \cL$.

\begin{figure}[t]
\centering
\includegraphics[width=8cm]{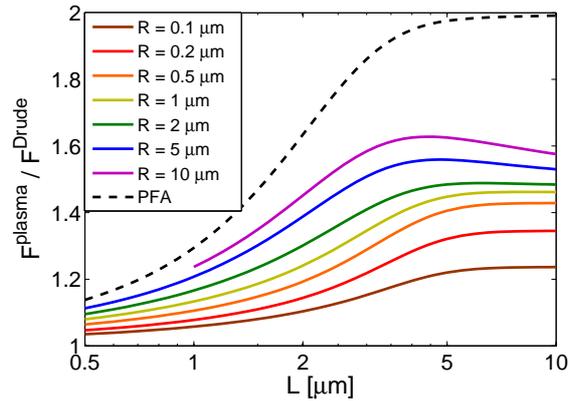}
\caption{Ratio of thermal Casimir force at $T=300K$
calculated with the plasma model and the Drude model,
as a function of surface separation $L$ for different
radii of the sphere.
The solid curves from bottom to top correspond
to increasing values of sphere radii.
The dashed curve is the PFA prediction.}
\label{plane-sphere-T-fig3}
\end{figure}

In the case of the Drude model with a non vanishing relaxation frequency
the free energy for the Drude model turns out to be 2/3 of the
expression for perfect mirrors whereas this ratio is 1/2 in the
plane-plane geometry.  The latter result is explained by the fact
that the TE reflection coefficient vanishes at zero frequency so
that only the TM modes contribute \cite{Bostrom00,BrevikNJP06}.
The change of the ratio 1/2 to 2/3 in the plane-sphere geometry has to
be attributed to the redistribution of the TE and TM contributions
into electric and magnetic spherical eigenmodes.
The change is illustrated in Fig. \ref{plane-sphere-T-fig3}, where we have plotted
the ratio of the thermal Casimir force $F^\plas$ calculated with the plasma model
to the one $F^\Drud $ obtained with the Drude model.
Again, the plots correspond to $\lambda_\P=136$nm and
$\lambda_\gamma/\lambda_\P=250$.
The results of our calculations are shown by the solid curves with
the sphere radius increasing from bottom to top. The ratio
$F^\plas/F^\Drud$ varies in the plane-sphere geometry as a function
of the sphere radius, clearly demonstrating the strong
interplay between the effects of temperature, dissipation and
geometry. For large spheres ($R \gg \lambda_\P$), the ratio converges
to the value 3/2,  whereas it remains smaller for small spheres
(down to $~1.2$ for $R\sim100$nm). The dashed curve gives the
variation of the same ratio as calculated within the PFA which leads
to a factor 2 in the limits of large distances or high temperatures,
corresponding to the prediction in the parallel-plates geometry.
This factor 2 deduced within PFA is never
approached within the calculations performed in the plane-sphere
geometry.

\section{Conclusion}

In this paper we have reviewed the quantum and thermal Casimir interaction between parallel plates, corrugated surfaces and plane and spherical mirrors. To perform our calculations we have extensively used the scattering approach where the objects are characterized by scattering matrices. We have compared our results with predictions obtained within the PFA. When taking the diffraction of the electromagnetic field correctly into account, surprising features appear especially for the thermal Casimir force in the plane-sphere geometry, where the exact results differ substantially from predictions within the PFA.
While open problems are still waiting to be tackled, the whole set of presented results clearly illustrates the usefulness and practicality of the scattering approach in Casimir physics.

\section*{Acknowledgments}
The authors thank I. Cavero-Pelaez, D. Dalvit, G.L. Ingold, M.-T. Jaekel,
J. Lussange, P.A. Maia Neto, R. Messina, P. Monteiro,
I. Pirozenkho and V. Marachevsky for contributions and/or fruitful discussions, H. B. Chan for kindly providing the data of his experiment
and the ESF Research Networking Programme CASIMIR (www.casimir-network.com)
for providing excellent possibilities for discussions and exchange.
Financial support from the French Contract ANR-06-Nano-062
and from Capes-Cofecub are gratefully acknowledged.

\end{document}